\documentclass[twocolumn,floatfix,pra,showpacs,preprintnumbers,superscriptaddress,amsmath,amssymb,10pt,aps]{revtex4}

\usepackage{mathtools}
\usepackage{mathptmx}
\usepackage{subfigure}
\usepackage{psfrag,graphicx}
\usepackage{dcolumn}
\usepackage{amsmath,amssymb}
\usepackage{bm}
\usepackage{color}
\usepackage{latexsym}
\usepackage{epstopdf}
\usepackage{color}
\usepackage[english]{babel}
\usepackage{latexsym}
\usepackage{psfrag,graphicx}
\usepackage{subfigure}
\usepackage{amsmath}
\usepackage{amssymb}
\usepackage{amsfonts}
\usepackage{bm}
\usepackage{natbib}
\usepackage{epstopdf}
\usepackage{xcolor}
\usepackage{appendix}

\definecolor{mygrey}{gray}{0.35}
\definecolor{myblue}{rgb}{0.2,0.2,0.8}
\definecolor{myzard}{cmyk}{0,0,0.05,0}
\definecolor{mywhite}{rgb}{1,1,1}
\definecolor{mywhite}{rgb}{1,1,1}
\definecolor{myred}{rgb}{1,0.,0.3}

\usepackage[colorlinks=true,citecolor=myblue,linkcolor=myred]{hyperref}

\def\ba{\begin{align}}
\def\enda{\end{align}}
\def\bi{\begin{itemize}}
\def\ei{\end{itemize}}

\def\be{\begin{equation}}
\def\ee{\end{equation}}
\def\bea{\begin{eqnarray}}
\def\eea{\end{eqnarray}}
\def\bse{\begin{subequations}}
\def\ese{\end{subequations}}

\begin{document}
\title{--}
\title{Onset of Quantum Thermalization in Jahn-Teller model}
\def\correspondingauthor{\footnote{Corresponding author: pivanov@phys.uni-sofia.bg}}
\author{Yoana R. Chorbadzhiyska}
\affiliation{Center for Quantum Technologies, Department of Physics, St. Kliment Ohridski University of Sofia, James Bourchier 5 blvd, 1164 Sofia, Bulgaria}
\author{Peter A. Ivanov}
\affiliation{Center for Quantum Technologies, Department of Physics, St. Kliment Ohridski University of Sofia, James Bourchier 5 blvd, 1164 Sofia, Bulgaria}

\begin{abstract}
We investigate the onset of quantum thermalization in a system governed by the Jahn-Teller Hamiltonian which describes the interaction between a single spin and two bosonic modes. We find that the Jahn-Teller model exhibits a finite-size quantum phase transition between the normal phase and two types of super-radiant phase when the ratios of spin-level splitting to each of the two bosonic frequencies grow to infinity. We test the prediction of the Eigenstate Thermalization Hypothesis in the Jahn-Teller model. We show that the expectation value of the spin observable quickly approaches its long-time average value. We find that the distance between the diagonal ensemble average and the microcanonical ensemble average of the spin observable decreases with the effective thermodynamic parameter. Furthermore, we show that the mean-time fluctuations of the spin observable are small and are inversely proportional to the effective system dimension. 

\end{abstract}

\maketitle


\section{Introduction}
An isolated non-integrable quantum system prepared in an out-of-equilibrium state undergoes a process known as quantum thermalization \cite{Popescu2006,Rigol2008,Eisert2014,Alessio2016}. One of the most successful approaches for description of this intriguing quantum phenomenon is the Eigenstate Thermalization Hypothesis (ETH). It assumes that the expectation values of an observable calculated in the basis of eigenstates of the non-integrable Hamiltonian are equal to the average calculated with the microcanonical ensemble \cite{Deutsch1991,Deutsch2018,Srednicki1994,Srednicki1996}. The validity of the ETH has been studied in various quantum many-body systems by means of exact diagonalization \cite{Rigol2009,Rigol2012,Swan2019,Jansen2019,kim2014,Villasenor,Kirkova2023,Nation2018}. Moreover, there have been key experiments demonstrating the quantum thermalization in isolated systems using ultracold atoms \cite{Gring2012,Kaufman2016}, superconducting qubits \cite{Neill2016}, and trapped ions \cite{Smith2016,Clos2016,Kranzl2023}. Usually, the quantum thermalization in these systems is associated with complexity increase as it requires a large Hilbert space growing exponentially with the system size \cite{Beugeling2014}.

In this work we study the emergence of quantum thermalization in a \emph{small} system consisting solely of a single spin and two bosonic degrees of freedom, which is represented by Jahn-Teller (JT) model. The JT model describes electronic orbitals coupled to vibrational modes either in molecules or solids \cite{Englman1972,Bersuker2006}. The JT effect in such systems is related to a structural instability in electronically degenerate states of molecules, where electron-phonon interaction shifts the potential minima of the nuclei, which causes distortion of the molecular configuration. Here we show that the quantum-optical analog of the JT model undergoes a finite-size second-order quantum phase transition from normal to two types of super-radiant phases when the ratios of spin level-splitting to each of the two bosonic frequencies grow to infinity. Such a second-order quantum phase transition is associated with macroscopically excited bosonic state of one of the bosonic modes. We also find that the JT model exhibits a first-order quantum phase transition between two super-radiant phases where the bosonic order parameters exhibit jump at the critical point.

We study the onset of thermalization in the different quantum phases of the JT model. We focus first on the equivalence between diagonal ensemble average and microcanonical ensemble average of the spin observable. In the normal phase, the large spin-level splitting prevents the relaxation of the spin observable and its time-evolution shows fast oscillations around the initial value. In the super-radiant phase, the time-evolution of the observable quickly reaches its long-time average value, around which oscillates for all subsequential times. The amplitudes of the oscillation, however, decrease as the effective thermodynamic parameter increases. We also show that the difference between the diagonal ensemble average and microcanical ensemble average of the observable decreases with the effective thermodynamic parameter.  Furthermore, we investigate the matrix elements of our observable in the basis of eigenstates of the JT Hamiltonian. For the eigenstates with energy close to the energy of the initial state of the JT system, we find that each diagonal matrix element is approximately equal to the microcanonical average, which is in agreement with the weak ETH. 

Further, we focus on the equilibration aspect of quantum thermalization which assumes that the mean time-fluctuations around the time average value of an observable are small and decrease with the system size. We approach this question by considering the effective time-averaged dimension $d_{\rm eff}$ of the JT model, which is a measure for ergodicity of the system \cite{Linden2009}. In the normal phase the effective dimension is small which prevents the thermalization. In contrast, we find that the effective dimension in the super-radiant phase is large and increases with the spin-boson couplings, which leads to large bosonic Hilbert space. 
The large effective dimension indicates that a large set of eigenvectors of the JT Hamiltonian are involved in the dynamics of the system such that destructive interferences can lead to a suppression of the amplitudes of the time fluctuations on average. Moreover, for ergodic quantum systems which can be described by the random matrix ansatz, the mean time-fluctuations decrease as $d_{\rm eff}^{-1/2}$ \cite{Nation2018}. We study the scaling of the mean time-fluctuations in our system with the effective dimension. We find that in the super-radiant phase the mean amplitude of time-fluctuations is inversely proportional to the effective system dimension.

Finally, we consider the second-order R\'enyi entropy (RE) which measures the entanglement between the spin subsystem and the two bosonic modes. In the normal phase the RE is nearly zero which indicates that the subsystems are disentangled. In the super-radiant phase the RE quickly increases and reaches a level of saturation, where it remains for all subsequential times which is a signature for thermalization.

The article is organized as follows: In Sec. \ref{JTM} we show that the JT model undergoes a finite-size quantum phase transition in the effective thermodynamic limit which is given by the ratio between the spin-level splitting and the bosonic frequencies. We show that when the JT spin-boson coupling increases the system undergoes quantum phase transition from the normal to a super-radiant phase, where one of the bosonic modes is excited. The JT system also exhibits a first-order quantum phase transition between two super-radiant phases. In Sec. \ref{QTJT} we discuss the onset of quantum thermalization in the JT system. We show that the spin observable quickly reaches its long-time average value which is approximately equal to the microcanonical prediction. We also show that the mean time fluctuation of the spin observable decreases with the effective dimension of the system, which is in agreement with ETH. Finally, the conclusion is presented in Sec \ref{S}.

\section{Jahn-Teller Model}\label{JTM}

We consider a model consisting of a single spin-1/2 system with energy splitting $\Delta$ and a two-dimensional quantum oscillator with mass $m$ and frequencies $\omega_{a}$ and $\omega_{b}$, which interact via Jahn-Teller coupling \cite{Larson2008}. The Hamiltonian is given by 
\begin{eqnarray}
\hat{H}_{\rm JT}&=&\frac{\hat{p}^{2}_{a}}{2m}+\frac{\hat{p}^{2}_{b}}{2m}+\frac{m\omega^{2}_{a}\hat{r}_{a}^{2}}{2}
+\frac{m\omega^{2}_{b}\hat{r}_{b}^{2}}{2}+\frac{\Delta}{2}\sigma_{z}\notag\\
&&+\mu_{a}\hat{r}_{a}\sigma_{x}+\mu_{b}\hat{r}_{b}\sigma_{y}.\label{JT0}
\end{eqnarray}
Here $\hat{p}_{i}$ and $\hat{r}_{i}$ $(i=a,b)$ are the momentum and the position operators of the quantum oscillator, and $\sigma_{x,y,z}$ are the Pauli matrices. The JT interaction between the spin and the quantum oscillators is given by the last two terms in (\ref{JT0}), where $\mu_{a}$ and $\mu_{b}$ are the coupling strengths. We introduce a pair of creation and annihilation operators $\hat{a}^{\dag}$, $\hat{a}$ and $\hat{b}^{\dag}$, $\hat{b}$ for each oscillator by writing $\hat{p}_{a}=ip_{0a}(\hat{a}^{\dag}-\hat{a})$ and $\hat{r}_{a}=r_{0a}(\hat{a}^{\dag}+\hat{a})$, and respectively $\hat{p}_{b}=ip_{0b}(\hat{b}^{\dag}-\hat{b})$ and $\hat{r}_{b}=r_{0b}(\hat{b}^{\dag}+\hat{b})$ where $p_{0i}=\sqrt{m\omega_{i}/2}$ and $r_{0i}=1/\sqrt{2m\omega_{i}}$ (we set $\hbar=1$ throughout the manuscript). Then, the JT Hamiltonian becomes
\begin{equation}
\hat{H}_{\rm JT}=\omega_{a}\hat{a}^{\dag}\hat{a}+\omega_{b}\hat{b}^{\dag}\hat{b}+\frac{\Delta}{2}\sigma_{z}+g_{a}\sigma_{x}(\hat{a}^{\dag}+\hat{a})+g_{b}\sigma_{y}(\hat{b}^{\dag}+\hat{b}),\label{JT}
\end{equation}
where we have omitted the constant term. Hence, in the new representation the JT Hamiltonian describes dipolar interaction between a single spin and two bosonic modes with spin-boson couplings $g_{i}=\mu_{i}r_{0i}$.  The total Hilbert space is spanned in the basis $\{|s\rangle|n_{a},n_{b}\rangle\}$, where $|s\rangle$ ($s=\uparrow,\downarrow$) is the eigenstate of $\sigma_{z}$ and $|n_{i}\rangle$ is the Fock state of the bosonic mode with occupation number $n_{i}$.

The physical realization of JT coupling has been discussed in various quantum-optical platforms, including, for example, cavity QED system \cite{Larson2008}, Bose-Einstein condensate \cite{Larson2009}, and trapped ions \cite{Porras2012,Ivanov2013}. The JT model describes intriguing quantum phenomena such as ground-state entanglement \cite{Hines2004,Liberti2007} and creation of artificial non-Abelian magnetic fields \cite{Larson2009_1}. Recently, a conical intersection and geometric phase have been experimentally observed with trapped ion system \cite{Whitlow2023}. The artificial molecular system with Rydberg atoms confined in optical tweezer traps was proposed in \cite{Magoni2023}, where the JT effect can be observed which is associated with distortion of the artificial molecular configuration. The many-particle extension of the JT model describes collective effects such as quantum chaos \cite{Majernikova2011} and magnetic-structural phase transition \cite{Porras2012,Ivanov2013,Ivanov2014}.

For general non-equal couplings $g_{a}\neq g_{b}$ the JT Hamiltonian (\ref{JT}) commutes with the parity operator $\hat{\Pi}=e^{i\pi(\hat{a}^{\dag}\hat{a}+\hat{b}^{\dag}\hat{b}+\frac{1}{2}(1+\sigma_{z}))}$ which measures an even-odd parity of total excitation number. When either $g_{a}\neq 0$, $g_{b}=0$ or $g_{b}\neq 0$, $g_{a}=0$ the Hamiltonian (\ref{JT}) describes the well-known quantum Rabi model \cite{Xie2017, Larson2021}. In the full symmetric case where $\omega_{a}=\omega_{b}$ and $g_{a}= g_{b}$ the JT Hamiltonian becomes U(1) invariant. In the following we focus on $Z$ parity symmetry JT model. Since it possesses two bosonic and one spin degrees of freedom the parity symmetry is not sufficient for our model to be integrable.

\subsection{Finite-size quantum phase transition}
In this work we consider the limit when the ratios of level-splitting $\Delta$ to bosonic frequencies $\omega_{i}$ grows to infinity, $\eta_{i}=\Delta/\omega_{i}\rightarrow\infty$, which essentially play the role of an effective thermodynamic limit in our model. Such a limit was considered for the quantum Rabi model, which exhibits a finite-size quantum phase transition \cite{Ashhab2013,Bakemeier2012,Hwang2015}. Here we show that the JT model undergoes a second-order quantum phase transition between normal and super-radiant phase, and a first-order quantum phase transition between the two super-radiant phases, which is associated with the jump of the order parameter at the critical point.

\subsection{Normal Phase}
It is convenient to introduce dimensionless spin-boson couplings $\lambda_{i}=2g_{i}/\sqrt{\omega_{i}\Delta}$. The JT system is in a normal phase when $\lambda_{i}\leq 1$, which is characterized by zero mean-field bosonic excitations of the two bosonic modes and polarized spin along the $z$ axis, namely (see Appendix \ref{Normal_phase})
\begin{equation}
\lim_{\eta_{a}\rightarrow\infty}\frac{\langle \hat{a}^{\dag}\hat{a}\rangle_{G}}{\eta_{a}}=0,\quad\lim_{\eta_{b}\rightarrow\infty}\frac{\langle \hat{b}^{\dag}\hat{b}\rangle_{G}}{\eta_{b}}=0,\quad \langle\sigma_{z}\rangle_{G}=-1.
\end{equation}
The excitations are $\epsilon_{\rm np}=\omega_{a}\sqrt{1-\lambda^{2}_{a}}+\omega_{b}\sqrt{1-\lambda^{2}_{b}}$,  which are real for $\lambda_{i}\le 1$.

\subsection{Super-radiant phase}
The presence of two critical couplings makes the transition to super-radiant phase more complicated. When these couplings attain values, such that $\lambda_{a}>1$ and $\lambda_{b}<1$, or equivalently $\lambda_{a}<1$ and $\lambda_{b}>1$, the system undergoes a second-order quantum phase transition to a super-radiant phase where either one of the two bosonic modes is macroscopically excited and the spin state is rotated along the $y$ or $x$ axis. We have for $\lambda_{a}<1$ and $\lambda_{b}>1$
\begin{eqnarray}
&&\lim_{\eta_{a}\rightarrow\infty}\frac{\langle \hat{a}^{\dag}\hat{a}\rangle_{G}}{\eta_{a}}=0,\quad \lim_{\eta_{b}\rightarrow\infty}\frac{\langle \hat{b}^{\dag}\hat{b}\rangle_{G}}{\eta_{b}}=\frac{\lambda^{4}_{b}-1}{4\lambda^{2}_{b}},\notag\\
&&\langle\sigma_{z}\rangle_{G}=-\frac{1}{\lambda^{2}_{b}}\label{SF}
\end{eqnarray}
and similarly for $\lambda_{a}>1$ and $\lambda_{b}<1$ interchanging $b\Leftrightarrow a$. The excitations now are $\epsilon_{\rm sp}=\omega_{a}\sqrt{1-(\lambda_{a}/\lambda_{b})^{2}}+\omega_{b}\sqrt{1-\lambda_{b}^{-4}}$ which are positively defined for $\lambda_{b}>1$ and $\lambda_{a}\leq\lambda_{b}$.
In the following we refer to these quantum phases as $a$- and $b$-super-radiant phases when either $a$ or respectively $b$ bosonic mode is excited.  

\begin{figure} 
\includegraphics[width=0.48\textwidth]{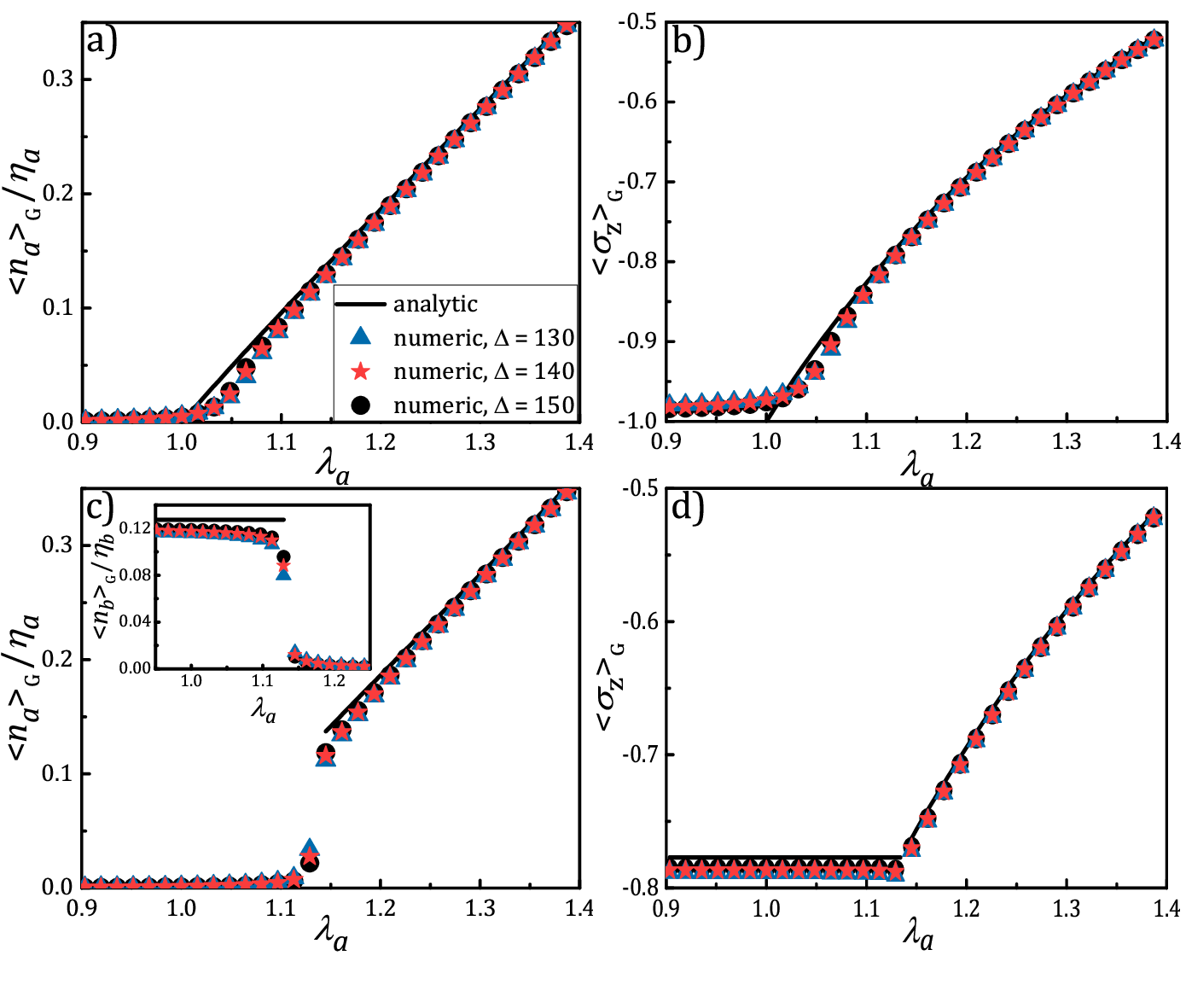}
\caption{Mean bosonic excitation $\langle\hat{n}_{a}\rangle_{G}/\eta_{a}$ and mean-field value of $\langle\sigma_{z}\rangle_{G}$ as a function of the dimensionless spin-boson coupling $\lambda_{a}$ for various $\Delta$. a) and b) Second order quantum phase transition between normal phase to $a$-super-radiand phase, where we set $\lambda_{b}=0.96$. c) and d) First order quantum phase transition between $b$-super-radiant phase and $a$-super-radiant phase. We set $\lambda_{b}=1.13$ and vary $\lambda_{a}$ across the transition point $\lambda_{a}=\lambda_{b}$. The mean bosonic excitations $\langle \hat{n}_{a}\rangle_{G}/\eta_{a}$ and $\langle \hat{n}_{b}\rangle_{G}/\eta_{b}$ (inset) show jump at $\lambda_{a}=\lambda_{b}$. The bosonic Hilbert space is truncated at $n_{\rm max}=80$ for $a$- and $b$-modes. }
\label{mean}
\end{figure}

Furthermore, we may consider the transition between $a$- and $b$-super-radiant phases by varying one of the spin-boson couplings. For instance, consider that $\lambda_{a}<1$ and $\lambda_{b}>1$ such that the system is in super-radiant phase where only $b$-mode is excited, see Eq. (\ref{SF}). Now, the system undergoes a first-order quantum phase transition to $a$-super-radiant phase when $\lambda_{a}>1$ and $\lambda_{a}>\lambda_{b}>1$. We have
\begin{eqnarray}
&&\lim_{\eta_{a}\rightarrow\infty}\frac{\langle \hat{a}^{\dag}\hat{a}\rangle_{G}}{\eta_{a}}=\frac{\lambda^{4}_{a}-1}{4\lambda^{2}_{a}},\quad \lim_{\eta_{b}\rightarrow\infty}\frac{\langle \hat{b}^{\dag}\hat{b}\rangle_{G}}{\eta_{b}}=0,\notag\\
&&\langle\sigma_{z}\rangle=-\frac{1}{\lambda^{2}_{a}}.
\end{eqnarray}
Since the quantum phase transition is associated with spontaneous symmetry breaking of the parity symmetry, only \emph{one} of the bosonic modes is excited. We plot in Fig. \ref{mean} the mean bosonic excitations and the mean field result for $\sigma_{z}$. In Figs. \ref{mean}(a) and \ref{mean}(b) we set $\lambda_{b}<1$ and vary the coupling $\lambda_{a}$. The JT system undergoes a second-order quantum phase transition from the normal phase to the $a$-super-radiant phase at critical coupling $\lambda_{a}(c)=1$. In Figs. \ref{mean}(c) and \ref{mean}(d) we set $\lambda_{b}>1$ such that the system is in a $b$-super-radiant phase and again vary $\lambda_{a}$. When $\lambda_{a}>\lambda_{b}$ the system undergoes a first-order quantum phase transition to $a$-super-radiant phase where the order parameters $\langle n_{i}\rangle_{G}$ exhibit discontinuity at the critical point $\lambda_{a}=\lambda_{b}>1$. 

\section{Onset of quantum thermalization in Jahn-Teller system}\label{QTJT}
In the following we investigate the onset of quantum thermalization in JT system in the different quantum phases. At the core of our understanding of thermalization of closed quantum system is the Eigenstate Thermalization Hypothesis which states that (i) expectation value of a thermalizing observable in the eigenstates of the non-integrable Hamiltonian is equal to the microcanonical prediction and (ii) the mean amplitudes of time-fluctuation decrease with the system size \cite{Deutsch1991,Deutsch2018,Srednicki1994,Srednicki1996}.

The energy spectrum of the JT Hamiltonian is $\hat{H}_{\rm JT}|\psi_{\mu}\rangle=E_{\mu}|\psi_{\mu}\rangle$, where $|\psi_{\mu}\rangle$ and $E_{\mu}$ are the eigenvectors and eigenergies. We assume that the system is initially prepared in an out-of-equilibrium state $|\Psi_{0}\rangle=\sum_{\mu}a_{\mu}|\psi_{\mu}\rangle$ with mean energy $E_{0}=\langle\Psi_{0}|\hat{H}_{\rm JT}|\Psi_{0}\rangle$ and this state evolves under the action of the unitary propagator $|\Psi(t)\rangle=e^{-i \hat{H}_{\rm JT}t}|\Psi_{0}\rangle=\sum_{\mu}a_{\mu}e^{-iE_{\mu}t}|\psi_{\mu}\rangle$. The long-time average of an observable $\hat{O}$ is given by
\begin{equation}
\langle \bar{O}\rangle=\lim_{\tau\rightarrow \infty}\frac{1}{\tau}\int_{0}^{\tau}\langle\Psi(t)|\hat{O}|\Psi(t)\rangle dt =\sum_{\mu}|a_{\mu}|^{2}O_{\mu\mu},
\end{equation}
where $O_{\mu\mu}=\langle\psi_{\mu}|\hat{O}|\psi_{\mu}\rangle$ and we have assumed that the eigenergies $E_{\mu}$ are non-degenerate. Hence, the long-time average of $\hat{O}$ can be written as $\langle \bar{O}\rangle={\rm Tr}(\hat{O}\hat{\rho}_{\rm DE})$, where $\hat{\rho}_{\rm DE}=\sum_{\mu}|a_{\mu}|^{2}|\psi_{\mu}\rangle\langle\psi_{\mu}|$ is the density matrix of the so-called diagonal ensemble (DE). The equilibration of a closed quantum system into a thermal state implies that
\begin{equation}
\langle \bar{O}\rangle\approx\langle O\rangle_{\rm micro},
\end{equation}
where $\langle O\rangle_{\rm micro}$ is the microcanonical average of $\hat{O}$ taken over an energy shell of eigenstates with energies $E_{\mu}$ close to $E_{0}$. 
\begin{figure} 
\includegraphics[width=0.48\textwidth]{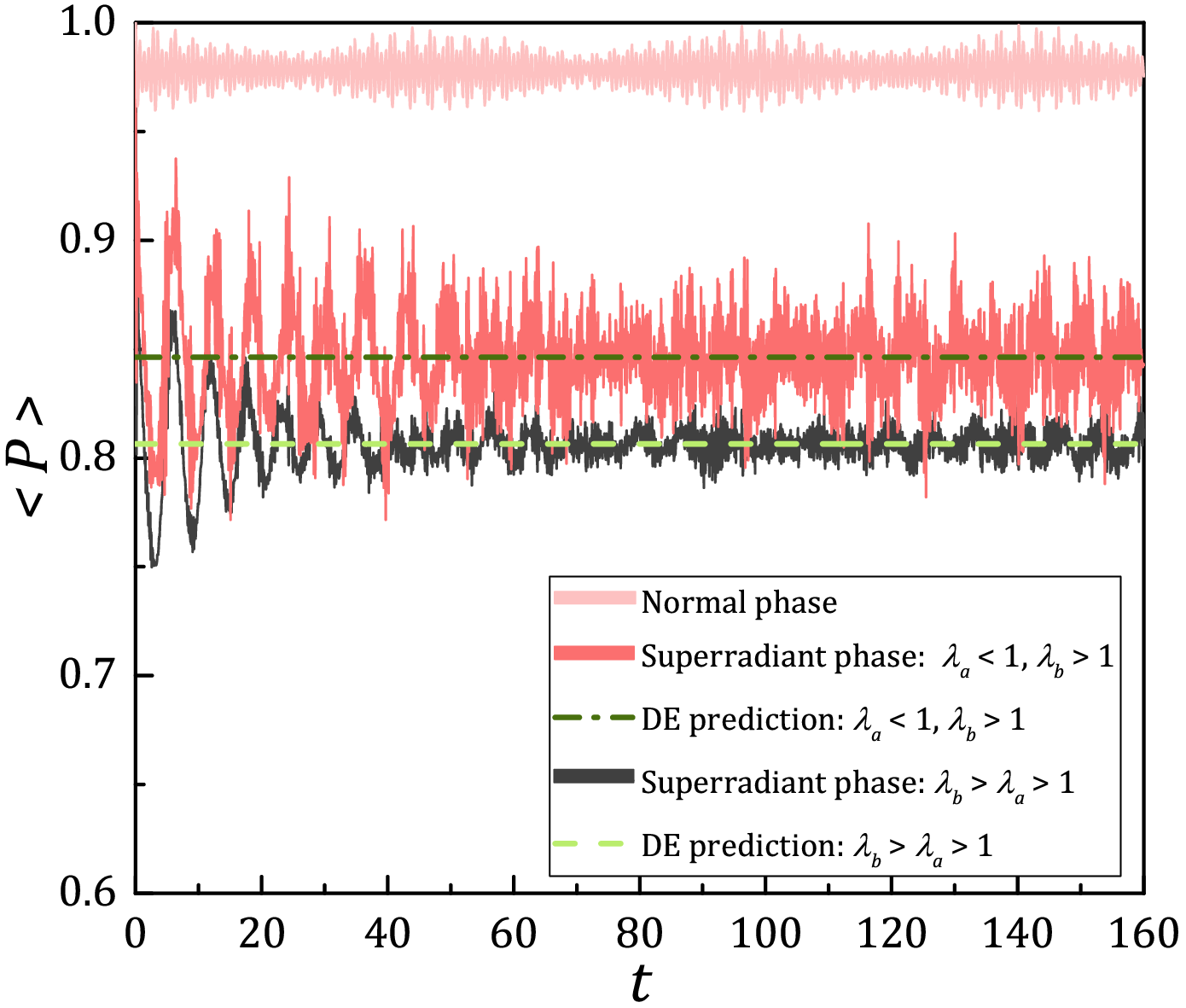}
\caption{Time-evolution of the spin population $\langle P(t)\rangle$ compared to the DE prediction in various quantum phases. The initial state is $|\Psi_0\rangle= \left|\downarrow\right\rangle|5_a,10_b\rangle$ and the parameters are set to: $\omega_a\Delta = 90$, $\omega_b\Delta = 100$ and $g_a=1.5$, $g_b=2.0$ (normal phase); $g_a=1.5$, $g_b=5.5$ ($\lambda_{a}<1$, $\lambda_{b}>1$); $g_a=5.2$, $g_b=5.5$ ($\lambda_{b}>\lambda_{a}>1$). }
\label{time_evol100}
\end{figure}
The ETH for the diagonal elements of an observable $\hat{O}$ suggests that the matrix element $O_{\mu\mu}$ is a smooth function of the energy $E_{\mu}$. The density matrix of the microcanonical ensemble is given by
\begin{equation}
\hat{\rho}_{\rm ME}=\frac{1}{\mathcal{N}}\sum_{\mu:|E_{\mu}-E_{0}|<\delta E}|\psi_{\mu}\rangle\langle\psi_{\mu}|,
\end{equation}
where the sum runs through the $\mathcal{N}$ eigenstates $|\psi_{\mu}\rangle$ of $\hat{H}_{\rm JT}$
that are within an energy shell of width $2\delta E$ centered at $E_{0}$. Hence, the microcanonical average of an observable $\hat{O}$ is $\langle O\rangle_{\rm micro}={\rm Tr}(\hat{O}\hat{\rho}_{\rm ME})$.

In Fig. \ref{time_evol100} we show the time evolution of the spin observable $\hat{O}=\hat{P}$, where $\hat{P}=\left|\downarrow\right\rangle\left\langle\downarrow\right|$ in the different quantum phases for initial state $|\Psi(0)\rangle=\left|\downarrow\right\rangle|n_{a},n_{b}\rangle$. In the normal phase, $\lambda_{i}<1$, the large energy splitting $\Delta$ suppresses the spin oscillation such that the spin population $\langle P(t)\rangle$ is nearly frozen around its initial value, namely $\langle P(t)\rangle\approx 1$. In the $b$-super-radiant phase $\lambda_{a}<1$ and $\lambda_{b}>1$ the spin population shows initial fast oscillations with large amplitude which subsequently decrease and tend to a long-time average value $\langle \bar{P}\rangle$. Again in the $b$-super-radiant phase, but now for $\lambda_{a}>1$ and $\lambda_{b}>1$, the spin population approaches faster the long-time average value, with smaller temporal fluctuations. 

\begin{figure} 
\includegraphics[width=0.48\textwidth]{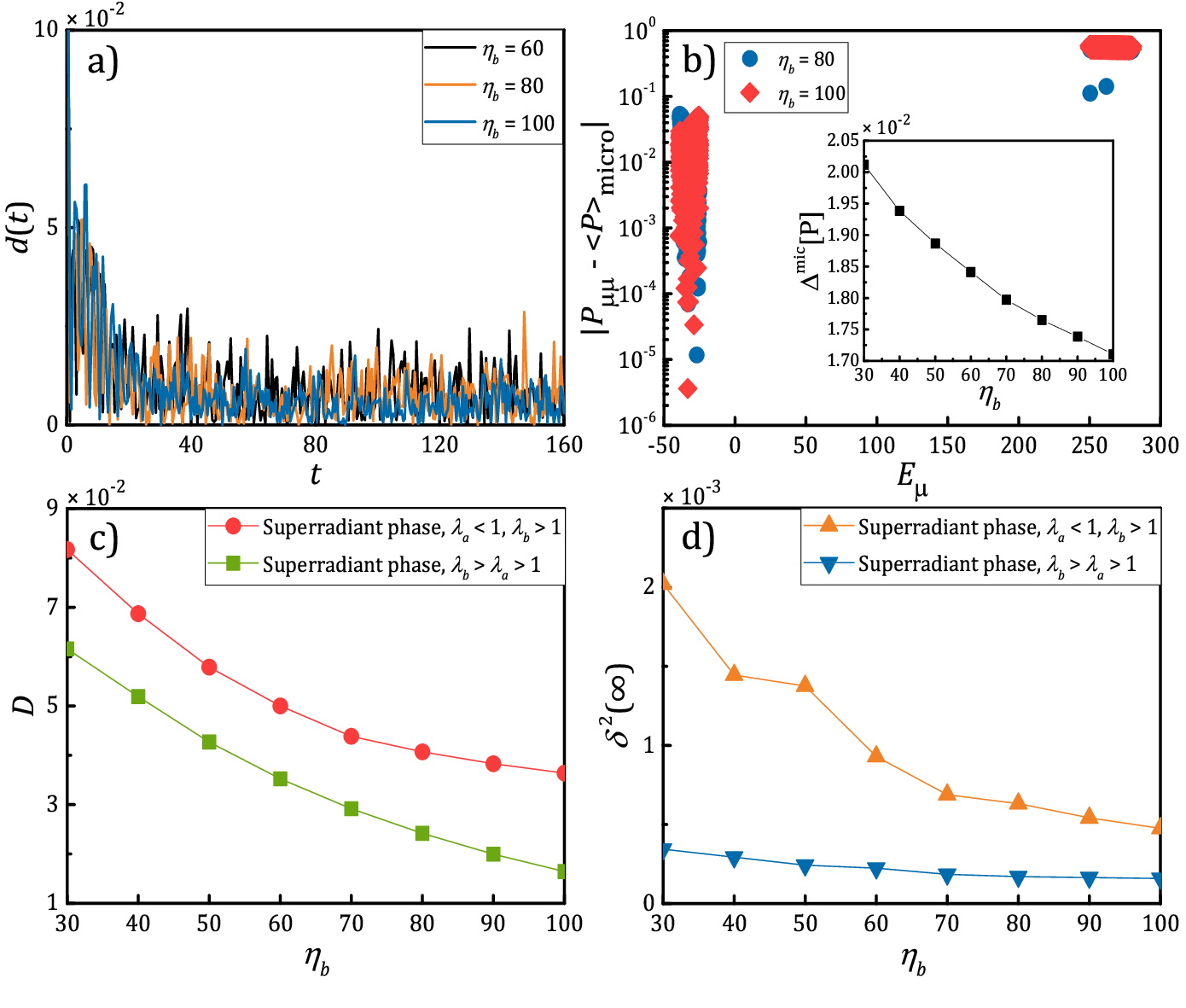}
\caption{a) The absolute difference between the expectation value of $\hat{P}$ and the average predicted by the DE for various $\eta_{b}$. We set $\lambda_{a}=1.096$ and $\lambda_{b}=1.10$ such that the system is in a $b$-super-radiant phase. b) The difference between the eigenstate expectation values and the microcanonical average in a $b$-super-radiant phase. (inset) Deviation of the EEVs for $\hat{P}$ with respect to the
microcanonical result as a function of the effective thermodynamic parameter. c) Comparison between the DE and ME averages of $\hat{P}$ as a function of the effective system size. d) The long-time fluctuations of the spin population. The bosonic Hilbert space is truncated at $n_{\rm max}=150$ for $a$- and $b$-modes. }
\label{3}
\end{figure}

In Fig. \ref{3}(a) we show the difference between the expectation value of our observable and its infinite time-average, $d(t)=|P(t)-\langle \bar{P}\rangle|$ as a function of time. We see that the expectation value quickly approaches the value predicted by the diagonal ensemble where it remains for all subsequential times, while oscillating with a small amplitude. The small time variations, however, decrease as we increase the effective thermodynamic parameter. 
We now compare our results with the microcanonical ensemble average. The ETH for diagonal elements of an observable consists of the assumption that the matrix element $O_{\mu\mu}$ is a smooth function of the energy $E_{\mu}$ such that microcanonical average is identical to the prediction of each eigenstate. In Fig. \ref{3}(b) we show $|P_{\mu\mu}-\langle P\rangle_{\rm micro}|$ for two regions of eigenstates of $\hat{H}_{\rm JT}$ as a function of the energy. We see that the expectation value for each eigenstate is approximately equal to the microcanonical average close to the mean energy and hence these eigenstates of the JT system exhibit thermal properties. However, for eigenstates with energies close to the edge of the JT spectrum we observe deviation between both results. The increase of the effective system size ameliorates this result. In order to show this, we consider the deviation of the eigenstate expectation values (EEVs) with respect to the microcanonical value \cite{Villasenor}
\begin{equation}
    \Delta^{\rm mic}[\hat{P}]=\frac{\sum_\mu|P_{\mu\mu}-\langle P\rangle_{\rm micro}|}{\sum_\mu P_{\mu\mu}},
\end{equation}
where the eigenstates are taken from an energy shell centered at the initial state energy. We observe monotonic decrease of the deviation with the effective thermodynamic parameter as is shown in Fig. \ref{3}(b) (inset).
Next, we study the difference $D=|\langle \bar{P}\rangle-\langle P\rangle_{\rm micro}|$ between the diagonal and microcanonical averages. In Fig. \ref{3}(c) we show the difference $D$ as a function of our effective thermodynamic parameter $\eta_{b}$ when the system is in a $b$-super-radiant phase for $\lambda_{a}<1$ and $\lambda_{b}>1$ as well as for $\lambda_{b}>\lambda_{a}>1$. In the limiting case $\lambda_{a}=0$ the JT system is reduced to quantum Rabi system. In that case we observe no agreement between both predictions \cite{Kirkova2022}. In the other two cases we see that the difference is small and decreases with $\eta_{b}$ where the agreement between both averages becomes more pronounced for $\lambda_{b}>\lambda_{a}>1$. Finally, we emphasize that the microcanonical energy shell is chosen to be $\delta E=p\Delta E$ where 
$\Delta E=E_{0}-E_{G}$ with $E_{G}$ being the ground state energy and $p$ varies from $25\%$ to $85\%$. The observed difference between the results for the microcanonical average is of order $10^{-3}$.

Further, we consider the long-time average of the temporal fluctuations of the expectation value of the observable $\hat{O}$ which is given by
\begin{equation}
\delta^{2}_{O}(\infty)=\lim_{\tau\rightarrow\infty}\frac{1}{\tau}\int_{0}^{\tau}\langle\Psi(t)|\hat{O}|\Psi(t)\rangle^{2}d\tau-\langle \bar{O}\rangle^{2}.
\end{equation}
Assuming non-degenerate energy gaps the infinite fluctuations can be expressed using the off-diagonal elements of an observable, namely $\delta^{2}_{O}(\infty)=\sum_{\mu\nu\atop\mu\neq\nu}|a_{\mu}|^{2}|a_{\nu}|^{2}|O_{\mu\nu}|^{2}$. In Fig. \ref{3}(d) we plot $\delta^{2}_{O}(\infty)$ for $\hat{O}={P}$ when the system is in a $b$-super-radiant phase. We see that the temporal fluctuations decrease with increasing $\eta_{b}$.
\begin{figure} 
\includegraphics[width=0.48\textwidth]{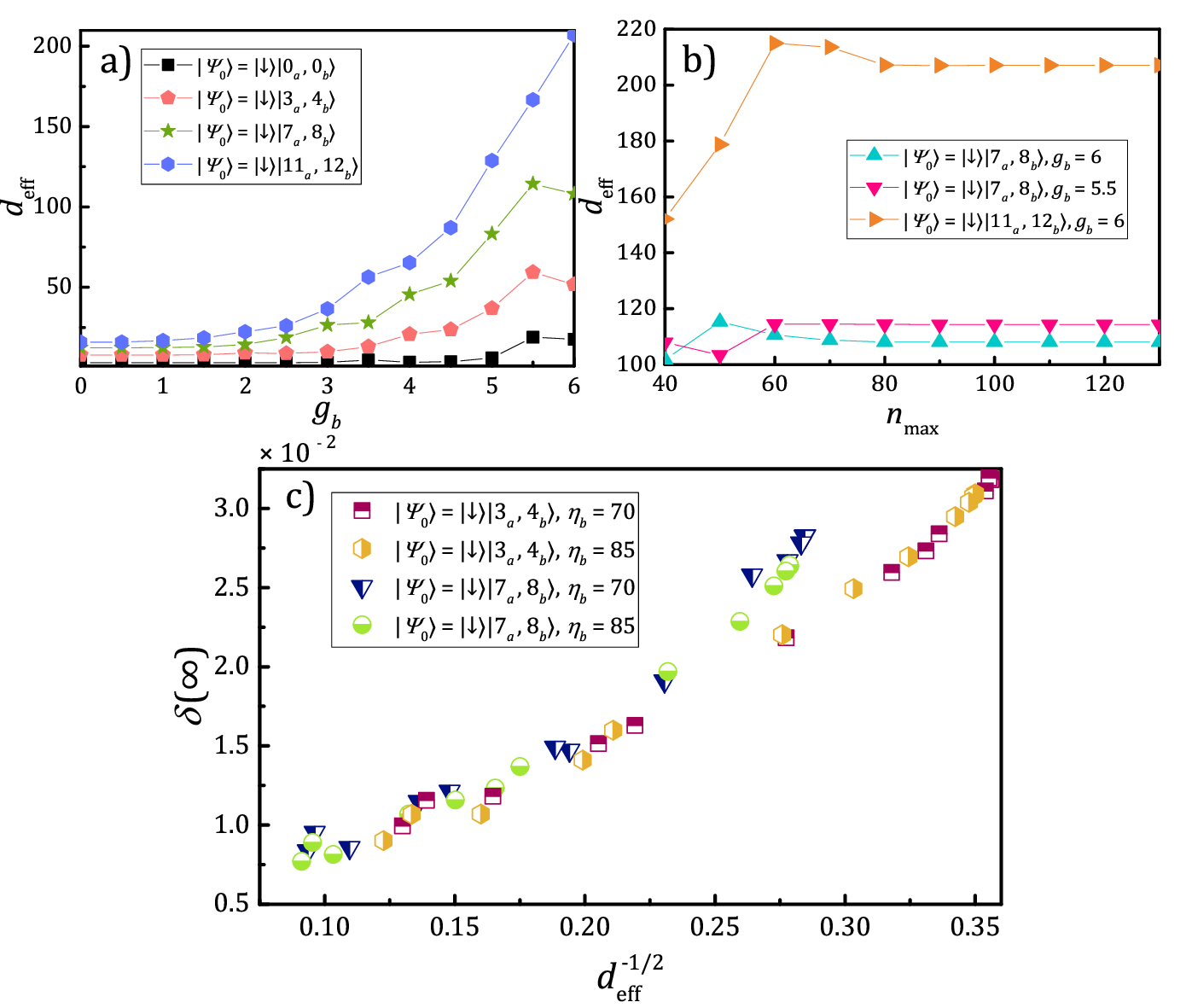}
\caption{a) The effective dimension as a function of the spin-boson coupling $g_{b}$ in $a$-super-radiant phase with $\lambda_{a}(g_{a}=5.9)=1.24$ and for various initial states. The other parameters are set to $n_{\rm max}=90$, $\eta_{a}=77.78$ and $\eta_{b}=70$. b) The effective dimension as a function of the bosonic Hilbert space truncation $n_{\rm max}$. c) The scaling of the long time-average of the temporal fluctuations of $\langle \bar{P}\rangle$ with the effective dimension $d_{\rm eff}$ in $a$-super-radiant phase. The bosonic Hilbert space is truncated at $n_{\rm max}=90$.}\label{4}
\end{figure}
As a measure for ergodicity of our system we introduce the effective dimension which is given by 
\begin{equation}
d_{\rm eff}=\frac{1}{\sum_{\mu}|c_{\mu}(\alpha)|^{4}}.
\end{equation}
Here we have assumed that the initial state is an eigenstate of non-interacting Hamiltonian $\hat{H}_{0}=\omega_{a}\hat{n}_{a}+\omega_{b}\hat{n}_{b}+(\Delta/2)\sigma_{z}$, namely $\hat{H}_{0}|\varphi_{\alpha}\rangle=E^{0}_{\alpha}|\varphi_{\alpha}\rangle$ such that $c_{\mu}(\alpha)=\langle\psi_{\mu}|\varphi_{\alpha}\rangle$. The effective dimension quantifies the ability of a quantum system to thermalize \cite{Linden2009}. Indeed, a large value of $d_{\rm eff}$ implies that a large set of eigenvectors are involved in the dynamics of an observable, such that destructive interferences  cause a suppression of the size of the time fluctuations on average. In Fig. \ref{4}(a) we show the effective dimension as a function of the spin-boson coupling $g_{b}$. We see that $d_{\rm eff}$ is large in the super-radiant phase and increases with the coupling strength. We also check the convergence of the result for $d_{\rm eff}$ by varying the bosonic Hilbert space truncation $n_{\rm max}$ for the two bosonic modes, see Fig. \ref{4}(b). We see that $d_{\rm eff}$ reaches a constant value for approximately $n_{\rm max}\approx 80$ for all cases. 

In Fig. \ref{4}(c) we show the scaling of the mean amplitudes of time-fluctuations with the effective dimension. Recently, it was shown that for systems which can be well described by a random matrix ansatz, the mean amplitudes of time-fluctuations scale as $\delta^{2}_{O}(\infty)\propto 1/d_{\rm eff}$ \cite{Nation2018}. We numerically check this scaling in our system. We vary $d_{\rm eff}$ through its dependence on the spin-boson coupling $g_{b}$. The long-time average is taken on sufficiently large time scales, which guarantees the convergence of the result. We find that the mean-time fluctuations follow approximately the inverse dependence of $d_{\rm eff}$ which is in agreement with ETH prediction.

\begin{figure} 
\includegraphics[width=0.48\textwidth]{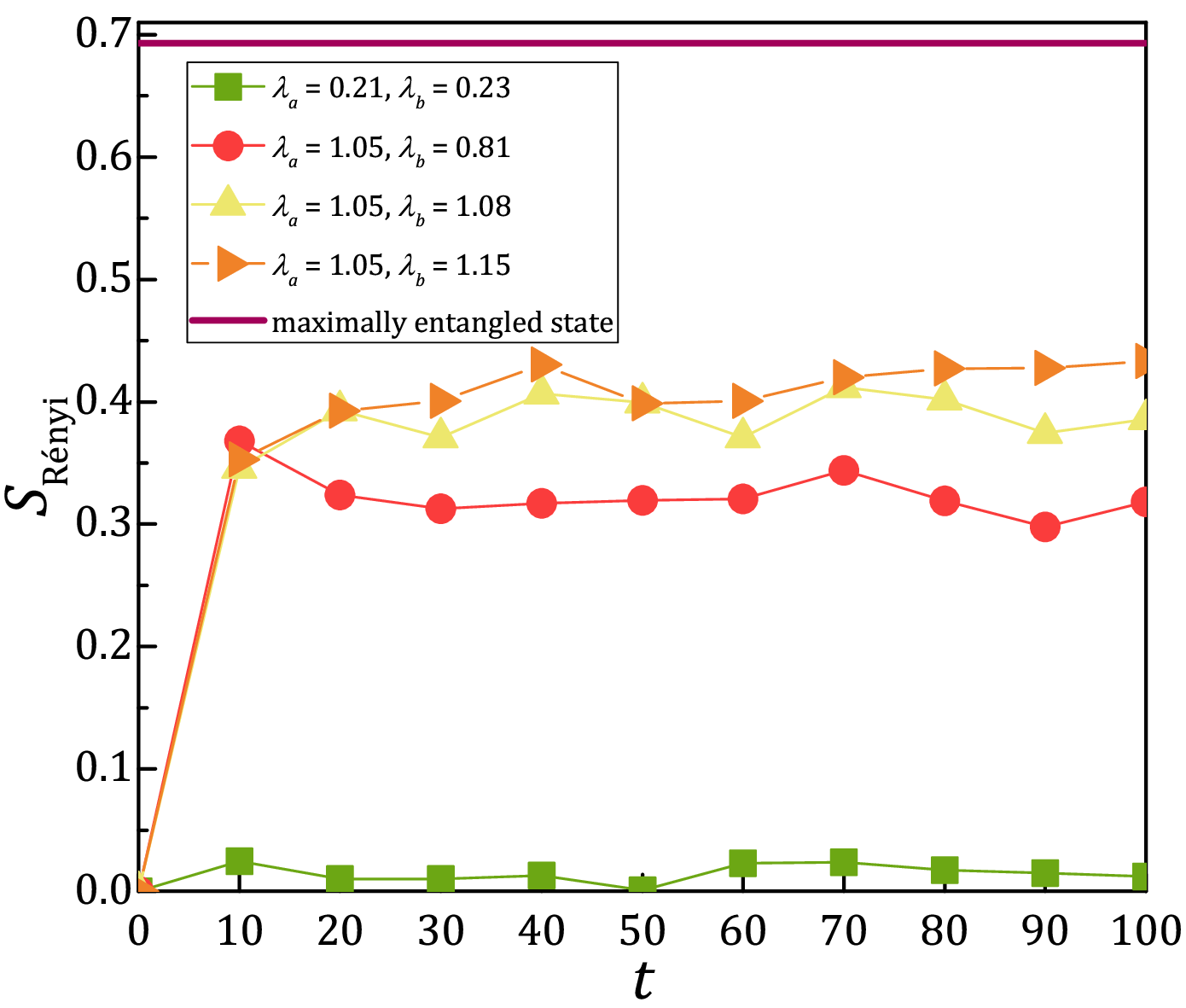}
\caption{a) The time evolution of the Rényi entropy in normal, $a$-super-radiant and $b$-super-radiant phase. The initial state is $|\Psi(0)\rangle=\left|\downarrow\right\rangle|7_{a},8_{b}\rangle$. The other parameters are set to $\omega_{a}\Delta=90$, $\eta_{b}= 75$ and $n_{\rm max}=90$.}\label{5}
\end{figure}

Finally, we study the Rényi entropy $S_{R}=-\ln {\rm Tr}(\hat{\rho}^{2}_{S})$ which measures the entanglement between the spin subsystem and its complement, where $\hat{\rho}_{S}$ is the reduced density matrix after tracing over the two bosonic degrees of freedom. As can be expected $S_{R}$ is nearly zero in the normal phase, which indicates that there is no entanglement between the spin system and the two bosonic modes, see Fig. \ref{5}. Increasing $\lambda_{i}$ the entanglement between the spin and the two bosonic modes increases as well. Accordingly the RE grows rapidly and reaches a level of saturation, which is signature for thermalization. The level of saturation increases with $\lambda_{i}$ and is more prominent for $\lambda_{i}>1$ as can be seen from Fig. \ref{5}.

\section{Summary}\label{S}
We have shown that the Jahn-Teller system consisting of single spin and two bosonic modes exhibits signatures of quantum thermalization. We have considered an effective thermodynamic limit in which the ratio of spin-level splitting to the bosonic mode frequencies grows to infinity. In this limit, the JT system undergoes a finite-size quantum phase transition between normal phase and two types of super-radiant phases. We have found that the JT undergoes a first-order quantum phase transition between two super-radiant phases where the order parameters show a jump at the critical point. We studied the signatures of quantum thermalization in the different quantum phases of JT system. In the normal phase the large energy-level splitting suppresses the time-evolution of the spin observable which is nearly frozen around its initial value. In the super-radiant phase, we found that the JT system quickly approaches its long-time average value. We have shown that the long-time average of the observable approaches the microcanonical ensemble average when the effective thermodynamic parameter is increased. Moreover, the individual matrix elements of the spin observable, corresponding to energies close to the initial one, are approximately equal to the microcanonical ensemble average which is in agreement with the weak ETH.

Further, we have shown that the mean time-fluctuation of an observable is small and decreases with the effective thermodynamic parameter. We also have studied the scaling of the mean time-fluctuation with the effective dimension of the JT system. According the ETH the mean time-fluctuation scales as $\delta_{O}^{2}(\infty)\propto 1/d_{\rm eff}$. We vary the effective dimension through its dependence on the JT spin-boson coupling and confirm that the mean time-fluctuation decreases with the effective dimension. Finally, we have shown the the Rényi entropy increases in the super-radiant phase and reaches a level of saturation which is signature for thermalization of the spin observable.

\section*{Acknowledgments}
This research is supported by the Bulgarian national plan for recovery and resilience, contract BG-RRP-2.004-0008-C01 (SUMMIT: Sofia University Marking Momentum for Innovation and Technological Transfer), project number 3.1.4. Y. R. C. acknowledges support by the SUMMIT, contract BG-RRP-2.004-0008, activity 3.4, project number No. 70-123-264. The authors acknowledge the use of the computational resources provided by the PHYSON HPC cluster and B. Milanov responsible for the smooth running of the PHYSON. 

\appendix

\begin{section}{Normal phase}\label{Normal_phase}
We consider the quantum Jahn-Teller Hamiltonian
\begin{equation}
\hat{H}_{\rm JT}=\omega_{a}\hat{a}^{\dag}\hat{a}+\omega_{b}\hat{b}^{\dag}\hat{b}+\frac{\Delta}{2}\sigma_{z}+g_{a}\sigma_{x}(\hat{a}^{\dag}+\hat{a})+g_{b}\sigma_{y}(\hat{b}^{\dag}+\hat{b}), \label{JT_A}
\end{equation}
in the effective thermodynamic limit $\eta_{i}\rightarrow\infty$. In order to describe the normal phase we perform canonical transformation $\hat{H}_1=e^{-\hat{A}_{\rm np}}\hat{H}_{\rm JT}e^{\hat{A}_{\rm np}}$ with $\hat{A}_{\rm np}^{\dag}=-\hat{A}_{\rm np}$. Applying the Baker–Campbell–Hausdorff formula to  $\hat{H}_1$ we obtain
\begin{eqnarray}
\hat{H}_1&=&\hat{H}_0 + \hat{H}_I+[\hat{H}_0,\hat{A}_{\rm np}]+[\hat{H}_I,\hat{A}_{\rm np}] + \frac{1}{2!}[[\hat{H}_0,\hat{A}_{\rm np}],\hat{A}_{\rm np}]\notag\\
&&+\frac{1}{2!}[[\hat{H}_I,\hat{A}_{\rm np}],\hat{A}_{\rm np}]+\dots,
\end{eqnarray}
where $\hat{H}_0=\omega_{a}\hat{a}^{\dag}\hat{a}+\omega_{b}\hat{b}^{\dag}\hat{b}+\frac{\Delta}{2}\sigma_{z}$ and $\hat{H}_I=g_{a}\sigma_{x}(\hat{a}^{\dag}+\hat{a})+g_{b}\sigma_{y}(\hat{b}^{\dag}+\hat{b})$.
Our goal is to eliminate the terms in $\hat{H}_1$ which are linear in either of the couplings $g_a$ or $g_b$. In order to achieve this we choose $\hat{A}_{\rm np}$ to satisfies $\hat{H}_I = - \left[\hat{H}_0,\hat{A}_{\rm np}\right]$. We find
\begin{equation}
\hat{A}_{\rm np} =-i\frac{g_a}{\Delta}\sigma_{y}(\hat{a}^{\dag}+\hat{a}) + i\frac{g_b}{\Delta}\sigma_{x}(\hat{b}^{\dag}+\hat{b}).
\end{equation}
In this limit the effective Hamiltonian is given by $\hat{H}_1 \approx \hat{H}_0 + \frac{1}{2}\left[\hat{H}_I,\hat{A}_{\rm np}\right]$. Hence, we obtain 
\begin{equation}
\hat{H}_1 = \omega_{a}\hat{a}^{\dag}\hat{a} + \omega_{b}\hat{b}^{\dag}\hat{b} + \frac{\Delta}{2}\sigma_{z} + \frac{g_a^2}{\Delta}\sigma_{z}(\hat{a}^{\dag}+\hat{a})^2 + \frac{g_b^2}{\Delta}\sigma_{z}(\hat{b}^{\dag}+\hat{b})^2 ,
\end{equation}
which is diagonal in the basis of $\sigma_{z}$. Consider the subspace corresponding to spin state $\left|\downarrow\right\rangle$ we get
\begin{eqnarray}
\hat{H}^{\downarrow}_1&=&\omega_{a}\left(1-\frac{2g_a^2}{\omega_{a}\Delta}\right)\hat{a}^{\dag}\hat{a} + \omega_{b}\left(1-\frac{2g_b^2}{\omega_{b}\Delta}\right)\hat{b}^{\dag}\hat{b}\notag\\
&&-\frac{g_a^2}{\Delta}({{}\hat{a}^{\dag}}^2+\hat{a}^2)-\frac{g_b^2}{\Delta}({{}\hat{b}^{\dag}}^2+\hat{b}^2)-\frac{\Delta}{2}-\frac{g^{2}_{a}}{\Delta}-\frac{g^{2}_{b}}{\Delta}. 
\end{eqnarray}
We apply the transformation $\hat{H}^{\downarrow}_{\rm np}=\hat{S}(r_a)\hat{S}(r_b)\hat{H}^{\downarrow}_1\hat{S}^{\dag}(r_a)\hat{S}^{\dag}(r_b)$ with $\hat{S}(r_{a})=e^{\frac{r_{a}}{2}(\hat{a}^{\dag 2}-\hat{a})}$ and $\hat{S}(r_{b})=e^{\frac{r_{b}}{2}(\hat{b}^{\dag 2}-\hat{b})}$ where the squeezing magnitude is given by 

\begin{equation}
r_{i}=-\frac{1}{4}\ln(1-\lambda^{2}_{i}).
\end{equation}
The final diagonal Hamiltonian is given by
\begin{equation}
\hat{H}^{\downarrow}_{\rm np}=\omega_{a}\sqrt{1-\lambda^{2}_{a}}\hat{a}^{\dag}\hat{a} + \omega_{b}\sqrt{1-\lambda^{2}_{b}}\hat{b}^{\dag}\hat{b} + E_{\rm np},
\end{equation}
where we define $\lambda_{i}=g_{i}/g_{i}(c)$ with $g_{i}(c)=\sqrt{\Delta\omega_{i}}/2$ being the critical coupling parameters. The normal phase is defined by $g_{\chi}<g_{\chi}(c)$ or equivalently $\lambda_{\chi}<1$. The ground state energy is
\begin{equation}
E_{\rm np}=-\frac{1}{2}(\Delta+\omega_{a}+\omega_{b})+\frac{\omega_{a}}{2}\sqrt{1-\lambda^{2}_{a}}
+\frac{\omega_{b}}{2}\sqrt{1-\lambda^{2}_{b}}.
\end{equation}
with corresponding ground state $|\Psi_{\rm np}\rangle_{G} = \hat{S}(r_a)\hat{S}(r_b)\left|\downarrow\right\rangle|0_a,0_{b}\rangle$. 

In the thermodynamic limit the normal phase is characterized with mean spin magnetization  $\langle\sigma_{z}\rangle_{G}=-1$ along the $z$-axis and mean bosonic excitation $\langle\hat{a}^{\dag}\hat{a}\rangle_G/\eta_a\rightarrow 0$, $\langle\hat{b}^{\dag}\hat{b}\rangle_G/\eta_b\rightarrow 0$.
\\
\end{section}

\begin{section}{Super-radiant phase}\label{Super_radiant_phase}

In order to describe the super-radiant phase when for example $\lambda_{a} > 1$ and $\lambda_{b} < 1$, we displace first the $a$-bosonic mode in the following way: The JT Hamiltonian $\hat{H}_{\rm JT}$ is given by \eqref{JT} and we now apply the displacement transformation $\hat{H}_{2}=\hat{D}^\dag(\alpha_{a})\hat{H}_{\rm JT}\hat{D}(\alpha_{a})$ with $\hat{D}(\alpha_{a})=e^{\alpha_{a}(\hat{a}^{\dag}-\hat{a})}$ where $\alpha_{a}$ is the displacement amplitudes. Thus we obtain
\begin{eqnarray}
\hat{H}_{2}&=&\omega_{a}(\hat{a}^{\dag}+\alpha_{a})(\hat{a}+\alpha_{a})+\omega_{b}\hat{b}^{\dag}\hat{b}+\frac{\Delta}{2}\sigma_{z}\notag\\
&&+g_{a}\sigma_{x}(\hat{a}^{\dag}+\hat{a}+2\alpha_{a})+g_{b}\sigma_{y}(\hat{b}^{\dag}+\hat{b}).  \label{JT_B}
\end{eqnarray}

Extracting solely non-interacting terms that characterize the spin, we have 
\begin{equation}
\hat{H}_{\rm spin} = \frac{\Delta}{2}\sigma_{z} + 2\alpha_{a}g_{a}\sigma_{x},
\end{equation}
This is a two-state Hamiltonian which has eigenenergies $\pm\tilde{\Omega}$, where $\tilde{\Omega}=\sqrt{\Delta^2+16\alpha_{a}^2g_{a}^2}$. The corresponding eigenvectors are 
\begin{equation}
|\tilde{\uparrow}\rangle = \sin(\theta)\left|\uparrow\right\rangle + \cos(\theta)\left|\downarrow\right\rangle,\quad
|\tilde{\downarrow}\rangle =\sin(\theta)\left|\downarrow\right\rangle -\cos(\theta)\left|\uparrow\right\rangle,        
\end{equation}
where the angle $\theta$ is given by $\cos(2\theta)=-\Delta/\tilde{\Omega}$. In the eigenstate basis the spin Hamiltonian is $\hat{H}_{\rm spin}=\frac{\tilde{\Omega}}{2}\tilde{\sigma_{z}}$ The Hamiltonian \eqref{JT_B} takes the form
\begin{eqnarray}
\hat{H}_{2}&=&\omega_{a}\hat{a}^\dag\hat{a} + \omega_{b}\hat{b}^\dag\hat{b} + \frac{\tilde{\Omega}}{2}\tilde{\sigma}_{z} + \left[\omega_{a}\alpha_{a}+g_{a}\sin(2\theta)\tilde{\sigma}_{z}\right]\notag\\
&&\times(\hat{a}^\dag+\hat{a})-\cos(2\theta)g_{a}\tilde{\sigma}_{x}(\hat{a}^\dag+\hat{a}) + g_{b}\tilde{\sigma}_{y}(\hat{b}^\dag+\hat{b})\notag\\ 
&&+\omega_{a}\alpha_{a}^2\label{JT_B_basis}.
\end{eqnarray}
In order to determine the displacement amplitude $\alpha_{a}$ we project the term which multiple $(\hat{a}^{\dag}+\hat{a})$ in (\ref{JT_B_basis}) onto the subspace corresponding to spin state $|\tilde{\downarrow}\rangle$. In this representation the condition for cancelling linear term in the bosonic operators leads to the following result for the displacement parameter:  
\begin{equation}
    \alpha_{a}^2 = \frac{\eta_{a}}{4\lambda_{a}^2}(\lambda_{a}^4-1).
\end{equation}
The Hamiltonian \eqref{JT_B_basis} becomes
\begin{eqnarray}
\hat{H}_{2}&=&\omega_{a}\hat{a}^\dag\hat{a} + \omega_{b}\hat{b}^\dag\hat{b} + \frac{\tilde{\Omega}}{2}\tilde{\sigma}_{z}+\tilde{g}_{a}\tilde{\sigma}_{x}(\hat{a}^\dag+\hat{a})+g_{b}\tilde{\sigma}_{y}(\hat{b}^\dag+\hat{b})\notag\\
&&+\omega_{a}\frac{\eta_{a}}{4\lambda_{a}^2}(\lambda_{a}^4-1),
\end{eqnarray}
where we set $\tilde{g}_{a}=-g_{a}\cos(2\theta)$.
Next, we perform transformation according to $\hat{H}_3=e^{-\hat{A}_{\rm sp}}\hat{H}_{2}e^{\hat{A}_{\rm sp}}$, where
\begin{equation}
\hat{A}_{\rm sp} =-i\frac{\tilde{g}_a}{\tilde{\Omega}}\tilde{\sigma}_{y}(\hat{a}^{\dag}+\hat{a})+i\frac{g_b}{\tilde{\Omega}}\tilde{\sigma}_{x}(\hat{b}^{\dag}+\hat{b}).
\end{equation}
Then, in the limit $\tilde{\Omega}\gg \omega_{i}$ the effective Hamiltonian becomes
\begin{eqnarray}
\hat{H}_{3}&=&\omega_{a}\hat{a}^\dag\hat{a} + \omega_{b}\hat{b}^\dag\hat{b} + \frac{\tilde{\Omega}}{2}\tilde{\sigma}_{z} + \frac{\tilde{g}_{a}^2}{\tilde{\Omega}}\tilde{\sigma}_{z}(\hat{a}^\dag+\hat{a})^2\notag\\
&&+\frac{g_{b}^2}{\tilde{\Omega}}\tilde{\sigma}_{z}(\hat{b}^\dag+\hat{b})^2+\omega_{a}\frac{\eta_{a}}{4\lambda_{a}^2}(\lambda_{a}^4-1).
\end{eqnarray} 
Projecting onto the subspace of spin state $|\tilde{\downarrow}\rangle$ and subsequent application of the squeeze operators $\hat{S}(\tilde{r}_{a})=e^{\tilde{r}_{a}(\hat{a}^{\dag 2}-\hat{a}^{2})/2}$ and $\hat{S}(\tilde{r}_{b})=e^{\tilde{r}_{b}(\hat{b}^{\dag 2}-\hat{b}^{2})/2}$ where
\begin{equation}
\tilde{r}_{a}=-\frac{1}{4}\ln\left(1-\frac{1}{\lambda^{4}_{a}}\right),\quad \tilde{r}_{a}=-\frac{1}{4}\ln\left(1-\frac{\lambda^{2}_{b}}{\lambda^{2}_{a}}\right),
\end{equation}
brings the Hamiltonian into diagonal form  
\begin{equation}
\hat{H}_{\rm a-sp}^{\downarrow} = \omega_{a}\sqrt{1-\frac{1}{\lambda_{a}^4}}\hat{a}^\dag\hat{a} + \omega_{b}\sqrt{1-\left(\frac{\lambda_{b}}{\lambda_{a}}\right)^2}\hat{b}^\dag\hat{b} + E_{a-\rm sp},
\end{equation}   
which is positively defined for $\lambda_{a}>1$ and $\lambda_{a}>\lambda_{b}$. The ground state energy is
\begin{eqnarray}
E_{a-\rm sp}&=&\frac{\Delta}{4\lambda_{a}^2}(\lambda_{a}^4-1) - \frac{\Delta\lambda_{a}^2}{2} - \frac{1}{2}(\omega_{a}+\omega_{b}) \notag\\
&&+ \frac{\omega_{a}}{2}\sqrt{1-\frac{1}{\lambda_{a}^4}}+\frac{\omega_{b}}{2}\sqrt{1-\frac{\lambda_{b}^2}{\lambda_{a}^2}}.
\end{eqnarray}
\end{section}
The corresponding ground state is displaced squeezed state for $a$-mode and squeezed state for $b$-mode, namely
\begin{equation}
|\Psi_{a-\rm sp}\rangle_{G}=\hat{S}(\tilde{r}_{a})\hat{D}(\alpha_{a})\hat{S}(\tilde{r}_{b})|\tilde{\downarrow}\rangle |0_{a},0_{b}\rangle.
\end{equation}
The $a$-super-radiant phase is characterized with mean spin orientation $\langle\sigma_{z}\rangle=-\lambda^{-2}_{a}$ and mean bosonic excitation $\lim_{\eta_{a}\rightarrow\infty}\langle \hat{a}^{\dag}\hat{a}\rangle_{G}/\eta_{a}=(\lambda^{4}_{a}-1)/4\lambda^{2}_{a}$ and $\lim_{\eta_{b}\rightarrow\infty}\langle \hat{b}^{\dag}\hat{b}\rangle_{G}/\eta_{b}=0$.

The JT system in the $a$-super-radiant phase may undergo a quantum phase transition to $b$-super-radiant phase, where $\lambda_{b}>1$ and $\lambda_{b}>\lambda_{a}$. In that case the quantum phase is characterized with $\langle\sigma_{z}\rangle=-\lambda^{-2}_{b}$ and mean bosonic excitation $\lim_{\eta_{b}\rightarrow\infty}\langle \hat{b}^{\dag}\hat{b}\rangle_{G}/\eta_{b}=(\lambda^{4}_{b}-1)/4\lambda^{2}_{b}$ and $\lim_{\eta_{a}\rightarrow\infty}\langle \hat{a}^{\dag}\hat{a}\rangle_{G}/\eta_{a}=0$.

\end{document}